\documentclass[twocolumn]{aastex61}
\usepackage{mathrsfs, mathtools}

\begin{document}

\title{Millimeter Spectral Indices and Dust Trapping By Planets in Brown Dwarf Disks}
\author{P.~Pinilla}
\correspondingauthor{Paola~Pinilla, Hubble fellow}
\affiliation{Department of Astronomy/Steward Observatory, The University of Arizona, 933 North Cherry Avenue, Tucson, AZ 85721, USA}
\email{pinilla@email.arizona.edu}

\author{L.~H.~Quiroga-Nu\~{n}ez}
\affiliation{Leiden Observatory, Leiden University, PO Box 9513, 2300 RA Leiden, The Netherlands.}
\affiliation{Joint Institute for VLBI ERIC (JIVE), Postbus 2, 7990 AA Dwingefloo, The Netherlands.}

\author{M.~Benisty}
\affiliation{Unidad Mixta Internacional Franco-Chilena de Astronom\'{i}a, CNRS/INSU UMI 3386 and Departamento de Astronom\'{i}a, Universidad de Chile, Casilla 36-D, Santiago, Chile.}
\affiliation{Univ. Grenoble Alpes, CNRS, IPAG, F-38000 Grenoble, France.}

\author{A.~Natta}
\affiliation{Dublin Institute for Advanced Studies, School of Cosmic Physics, 31 Fitzwilliam Place, Dublin 2, Ireland.}
\affiliation{INAF-Arcetri, Largo E. Fermi 5, I-50125 Firenze, Italy.}

\author{L.~Ricci}
\affiliation{Department of Physics and Astronomy, Rice University, 6100 Main Street, 77005 Houston, TX, USA.}

\author{Th.~Henning}
\affiliation{Max Planck Institute for Astronomy, K\"onigstuhl 17, D-69117 Heidelberg, Germany.}

\author{G.~van~der~Plas}
\affiliation{Univ. Grenoble Alpes, CNRS, IPAG, F-38000 Grenoble, France.}

\author{T.~Birnstiel}
\affiliation{University Observatory, Faculty of Physics, Ludwig-Maximilians-Universit\"at M\"unchen, Scheinerstr. 1, 81679 Munich, Germany.}

\author{L.~Testi}
\affiliation{European Southern Observatory, Karl-Schwarzschild-Str. 2, D85748 Garching, Germany}
\affiliation{INAF-Arcetri, Largo E. Fermi 5, I-50125 Firenze, Italy.}

\author{K.~Ward-Duong}
\affiliation{School of Earth and Space Exploration, Arizona State University, Tempe, AZ 85287, USA.}

\begin{abstract}
Disks around brown dwarfs (BDs) are excellent laboratories to study the first steps of planet formation in cold and low-mass disk conditions. The radial-drift velocities of dust particles in BD disks are higher than in disks around more massive stars. Therefore, BD disks are expected to be more depleted in millimeter-sized grains compared to disks around T Tauri or Herbig Ae/Be stars. However, recent millimeter observations of BD disks revealed low millimeter spectral indices, indicating the presence of large grains in these disks and challenging models of dust evolution. We present 3\,mm photometric observations carried out with the IRAM/Plateau de Bure Interferometer (PdBI) of three BD disks in the Taurus star forming region, which have been observed with ALMA at 0.89\,mm. The disks were not resolved and only one was detected with enough confidence ($\sim3.5\sigma$) with PdBI. Based on these observations, we obtain the values and lower limits of the spectral index and find low values ($\alpha_{\rm{mm}}\lesssim 3.0$). We compare these observations in the context of particle trapping by an embedded planet, a promising mechanism to explain the observational signatures in more massive and warmer disks. We find, however, that this model cannot reproduce the current millimeter observations for BD disks, and multiple-strong pressure bumps globally distributed in the disk remain as a favorable scenario to explain observations. Alternative possibilities are that the gas masses in BD disk are very low ($\sim2\times10^{-3}\,M_{\rm{Jup}}$) such that the millimeter grains are decoupled and do not drift, or fast growth of fluffy aggregates.  
 \end{abstract}


%%%%%%%%%%
 \section{Introduction}     \label{introduction}
%%%%%%%%%%

Disks around Brown Dwarfs (BDs) are excellent laboratories to study the first steps of planet formation in cold and low-mass disks. Substantial circumstellar material based on mid-infrared, far-infrared, and  (sub-)millimeter emission has been observed around young low-mass stars, including BDs \citep[e.g.][]{klein2003, luhman2007, pascucci2009, harvey2012a, harvey2012b, joergens2012, vanderplas2016}. These observations revealed that BD disks are potential sites of planet formation \citep[e.g.][]{apai2005}, or even around free-floating planets \citep{bayo2017}. Observations with Herschel of the [OI] 63\,$\mu$m line and (sub-)millimeter continuum observations with ALMA show that BD disks are smaller and much less massive than disks around T Tauri stars, implying that planet formation may be limited around BDs \citep{testi2016, hendler2017}. However, companions around BDs have also been observed, as in the case of 2MASSWJ\,1207334-393254, a BD with mass of $M=25\,M_{\mathrm{Jup}}$, and with a companion of $\sim$5-7\,$M_{\mathrm{Jup}}$ detected at $\sim$55\,au \citep{chauvin2004}. The high-mass ratio between this BD and the companion ($\gtrsim$0.2) may suggest that they formed as a binary system, either by collapse of molecular cloud cores with sub-stellar masses \citep{padoan2004} or by dynamical ejection when a dense and unstable molecular cloud fragments and forms multiple systems \citep{reipurth2001, bate2009, bate2012}.

Recent observations of BD disks at millimeter-wavelength suggest that even in these very low-mass and cold conditions,  micron-sized dust particles grow to large sizes and these pebbles remain in the disk for million-year timescales \citep[e.g.][]{ricci2012}. This conclusion is based on the slope of the spectral energy distribution (SED) at long wavelengths, which can be interpreted in terms of grain size \citep[e.g.][]{henning1995, beckwith2000, draine2006}, with low values of  the spectral index ($\alpha_{\mathrm{mm}}\lesssim 3$) corresponding to large grains. This interpretation is valid when the emission is optically thin and in the Rayleigh-Jeans regime of the spectrum. Assuming disk midplane temperatures of ($\sim$20-100\,K), at (sub-)millimeter wavelengths ($\gtrsim$ 0.4\,mm), the Rayleigh-Jeans regime is suitable.

The initial growth of particles in protoplanetary disks is governed by the interaction with the gas. Because of the sub-Keplerian motion of the gas, the dust particles experience a headwind that leads to the loss of angular momentum and their fast inward drift \citep{whipple1972,weidenschilling1977}. Theoretical models predict that the radial-drift barrier is a devastating problem for millimeter dust particles in disks around low-mass stars, in particular, BDs, compared to their high-mass counterparts. The inward drift velocity of particles depends on how different the gas azimuthal velocity is with respect to the Keplerian speed. Because this difference is higher for disks around low-mass stars, the inward drift velocity of particles can be twice as fast for particles in BD disks than in disks around Sun-like stars \citep[e.g.][]{pinilla2013}, depleting the millimeter dust in the whole disk in short timescales ($\lesssim1000\,$yr). In addition, there is an observational correlation between the disk dust mass and the stellar mass \citep[see, e.g. Fig.~6 in][]{vanderplas2016}, indicating that the disk mass scales with the mass of the central object. Hence, BD disks are also low-mass disks, implying that centimeter-sized dust particles in the inner disk (few astronomical units from the BD) start to be decoupled from the gas and experience high inward drift velocities. A recent ALMA survey of disks in the Chamaeleon\,I star-forming region hinted that the radial-drift timescales are indeed shorter in disks around lower-mass stars \citep{pascucci2016}. Millimeter observations of BD disks have challenged current dust evolution models and only under extreme conditions where the radial drift is highly reduced in the entire disk, dust evolution models can explain current millimeter observations of BD disks \citep[][]{pinilla2013}.

To overcome the radial-drift barrier and explain the presence of millimeter grains in typical protoplanetary disks, the existence of a single broad pressure bump or multiple pressure bumps distributed radially have been suggested \citep[e.g.][]{klahr1997}. The presence of such pressure bumps can lead to  bright ring-like structures observable at the optical, near-infrared, and (sub-)millimeter wavelengths, as have recently been observed \citep[e.g.][]{alma2015, andrews2016, boer2016, isella2016, ginski2016, fedele2017, boekel2017, vanderplas2017}. Embedded planets can open gaps in disks and also lead to particle trapping at the outer edge of the gap. This scenario (in particular, when a giant planet is embedded in the disk), can create structures as observed in transition disks, which are a set of disks that lacks emission at $\lambda\lesssim10\,\mu$m and have dust cavities \citep[e.g.][]{strom1989,espaillat2014}. However, alternative models, such as photoevaporation and magnetohydrodynamical processes, can also create structures like a transition disk \citep[e.g.][]{flock2015, owen2016, pinilla2016}. In fact, strong coronal mass ejections and flares \citep[such as the ones reported by][for an ultracool dwarf]{schmidt2014} can trigger X-ray emissions that are high enough to increase the disk heating and ionization. These X-rays can trigger photoevaporation \citep{owen2011}, creating inner holes. At the edge of these inner holes, dust particles can be trapped \citep{alexander2007}. On the other hand, if X-rays can penetrate the disk, this can change the ionization of the inner disk, which can change the disk turbulence. This can lead to an active inner disk followed by a dead zone, which can have a direct consequence in the gas surface density profile \citep{flock2016}, and on the dust dynamics and growth \citep{pinilla2016}. Nonetheless, the duration of X-ray flares is usually much shorter than the growth time of magnetohydrodynamical turbulence or  than the cavity-opening process by photoevaporation \citep{ilgner2006}.  In these two cases (photoevaporation or dead zones), a single and strong pressure bump is expected as in the case of a single massive planet.

\begin{table*}
\label{table_disks}    
\caption{Observed BD disks with PdBI}
\begin{tabular}{c||ccccccccc}       
\hline
\hline
{\small \textbf{Used}} & {\small R.A.} &  {\small Decl.} & {\small SpT} & {\small $M_\star$} &{\small $L_\star$} &{\small $F_{\mathrm{0.89\,mm}}\pm\sigma$}&  {\small $F_{\mathrm{3.0\,mm}} \pm\sigma$} & {\small $\alpha_{0.89-3\,\rm{mm}}$} &{\small $M_{\rm{disk}, \rm{dust}}$}  \\ 
{\small \textbf{ID}}&{\small [J2000]}&{\small [J2000]}&&{\small ($M_\odot$)}&{\small ($L_\odot$)}&{(mJy)}& {\small ($\mu$Jy)}&&{\small ($M_\oplus$)}\\
\hline
{\small J043814} & {\small 04 38 14.76} & {\small +26 11 41.24} &  {\small M7.25} & {\small 0.065} &{\small 0.3}& {\small 1.57$\pm$0.16}  & 140 $\pm$ 60$^*$ & $\gtrsim$ 2.0 $\pm$ 0.5 & {\small 0.41} \\
{\small J043903} & {\small 04 39 03.69} & {\small +25 44 26.52} & {\small M7.25} &{\small 0.065}& {\small 0.02} & {\small 2.28$\pm$0.15} & 164 $\pm$ 60$^*$&  $\gtrsim$ 2.1 $\pm$ 0.4 & {\small 1.23}\\
{\small J044148} & {\small 04 41 48.04} & {\small +25 34 32.42} & {\small M7.75} &{\small 0.040} &{\small 0.01} & {\small 3.52$\pm$0.16} & 235 $\pm$ 64 & 2.2 $\pm$ 0.3 & {\small 2.34} \\
\hline
\hline
\end{tabular} 
\tablecomments{Column 1: Used ID. Column 2 and 3: Right ascension and declination. Column 3: Spectral type (SpT). Column 4: BD mass. Column 5: BD luminosity used for calculating the dust temperatures and disk dust masses \citep[data from][]{luhman2004, scholz2006}. Column 6: Total fluxes at 0.89\,mm from ALMA observations. Column 7: Total fluxes at 3.0\,mm from PdBI observations (*\,indicated upper limits). Column 8: Spectral index based on the 0.89 and 3.0\,mm fluxes. The uncertainties of the spectral index include 10\% of calibration error. Column 9: Estimated disk dust mass according to Eq.~\ref{mm_dust_mass} and assuming the fluxes at 0.89\,mm.} 
\end{table*}

\begin{figure*}
 \centering
 \tabcolsep=0.05cm 
   \begin{tabular}{ccc}   
   	\includegraphics[width=6cm]{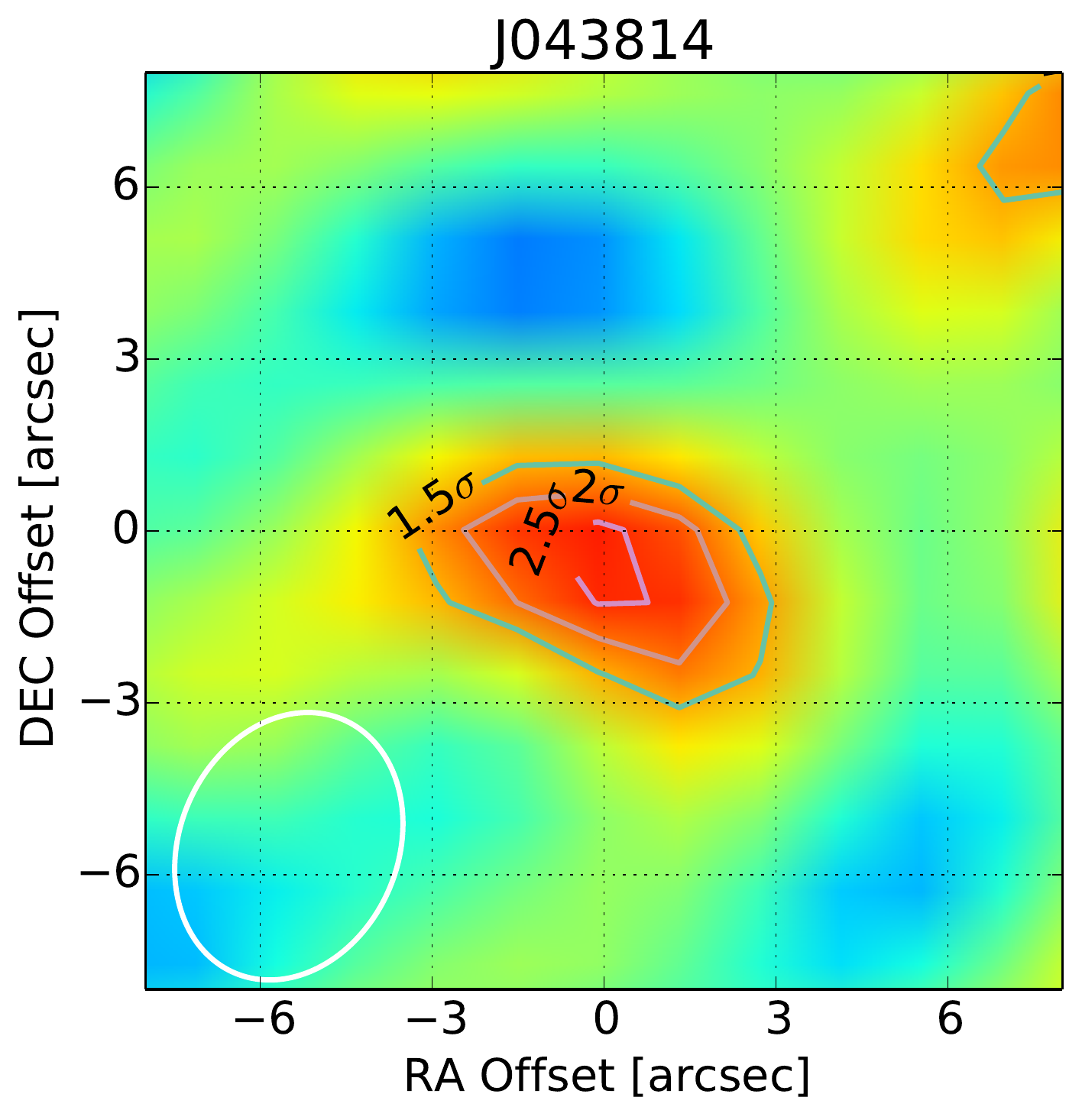}&
	\includegraphics[width=6cm]{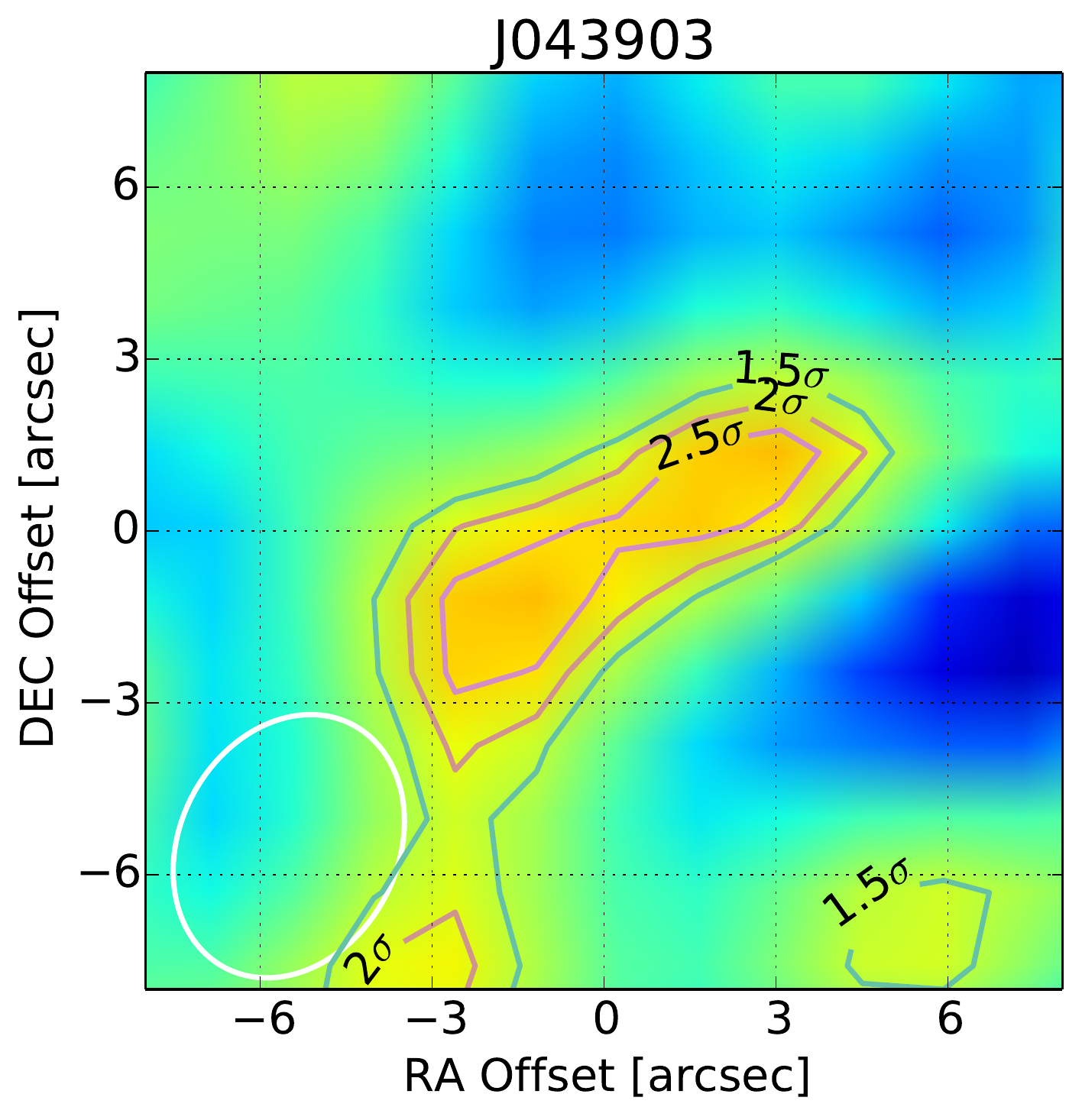}&
	\includegraphics[width=6cm]{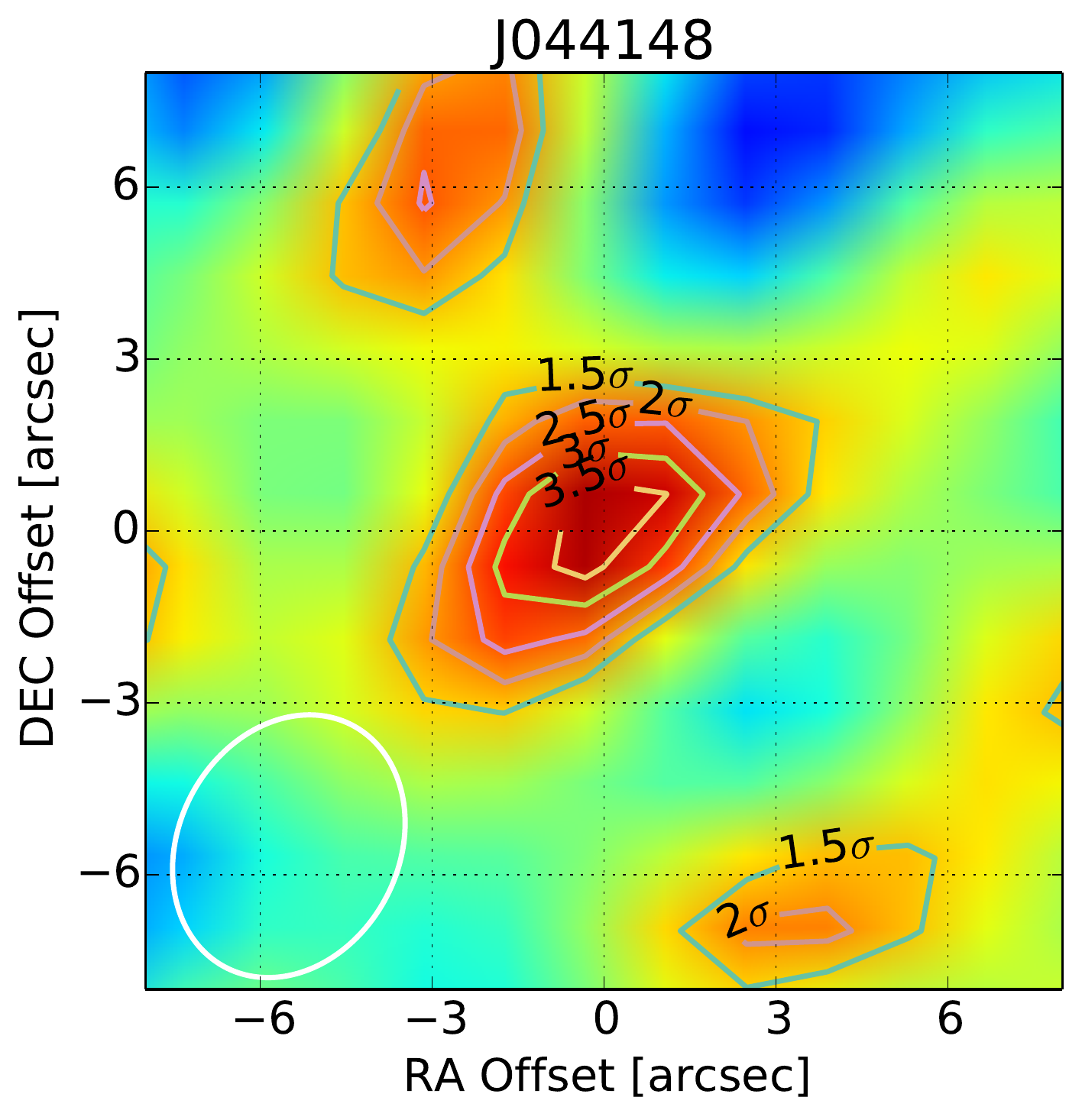}
   \end{tabular}
   \caption{Continuum maps at 3\,mm obtained with PdBI for three BD in Taurus (Table~\ref{table_disks}). The white circle in the left corner shows the size and position of the beam. Contours are plotted based on the flux information shown in Table~\ref{table_disks}.}
   \label{continuum_maps}
\end{figure*}

A couple of disks around BDs have been identified as transition disks in IC~348 \citep{muzerolle2006},  with small cavity sizes of few au, but larger than the magnetospheric truncation or dust sublimation radii. \cite{muzerolle2006} explained that the UV flux expected in BD disks is probably too low for the dispersal of the inner disk and to explain the cavity origin by photo-evaporation. Embedded planets that are massive enough to open a gap may be an option; however, \cite{payne2007} suggested that giant planet formation is inhibited in that sub-stellar regime and only Earth-like planets can form via core accretion. Hence, \cite{payne2007} concluded that Jupiter-like companions to BDs are only possible to form as binary systems. In addition, \cite{testi2016} systematically examined the disk masses of BD disks in Ophiuchus, and found that at 1\,Myr the reservoir of material available for planet formation is very small in BD disks. The obtained disk dust masses depend on the assumption for the dust temperature, as shown in Fig.~A1 from \cite{testi2016}. Nonetheless, independent of the assumption for the dust temperature, according to their observations, the BD disk masses are not sufficient to build Jupiter-mass planets, but also very small to build several Earth-like planets. 

In this paper, we explore whether dust trapping by a massive planet (which may not form in the disk) can lead to low millimeter spectral indices as observed in BD disks. In the case of disks around T Tauri or Herbig Ae/Be stars, these models of trapping by embedded planets are enough to keep millimeter grains in the outer regions on million-year timescales and to explain the millimeter spectral index, in particular, for transition disks \citep{pinilla2014}. 

In order to investigate trapping by a planet in a BD disk, we first aim to answer the question: What is the minimum mass planet needed to open a gap and trap particles in a BD disk? Afterwards, assuming such a planet, we explore particle trapping and whether or not the resulting dust density distributions at million-year timescales can reproduce low values of the spectral index.  We compare our models results with new values and limits of the spectral index obtained for three disks around BDs. We present new 3\,mm (100 GHz) photometric observations carried out with the IRAM/Plateau de Bure Interferometer (PdBI\footnote{Now the NOrthern Extended Millimeter Array, NOEMA.}) of three BD disks in Taurus, which were observed with ALMA in Band 7 ($\sim$0.89\,mm, 339\,GHz). With these two millimeter fluxes, we calculate the spectral index for each target, doubling the previous measurement of spectral index estimations for BD disks. 

This paper is organized as follows. In Sect~\ref{sect:observations}, we present a short description of the PdBI observations and the calculation of the disk dust masses. Moreover, we obtain the millimeter spectral index and compare these results to other star-formation regions. In Sect~\ref{sect:models}, we present our calculations  for a planet to open a gap and trap particles in a BD disk, we show the results of 2D hydrodynamical simulations and dust evolution models, together with the comparison with current observations of BD disks. In Sect.~\ref{sect:discussion} and~\ref{sect:conclusion}, we discuss the results and present the main conclusion of this work respectively. 
 
\section{Observations} \label{sect:observations}
The observed targets are located in the Taurus star-formation region \citep[at a distance of 140\,pc,][]{wichmann1998}, and they were previously identified by photometric surveys and confirmed as Taurus members by optical spectroscopy~\citep[e.g.][]{martin2001, briceno2002, luhman2003}. Based on the available flux information at 1.3\,mm \citep{scholz2006}, we selected the three disks with the highest millimeter fluxes (J043814, J043903, and J044148; see Table~\ref{table_disks}). 

We obtained 3.0\,mm photometric observations of three disks around BDs with PdBI.  The observations were made on 2014  July 23, 26, and 31. The most compact configuration of five antennas was used to have the maximum sensitivity possible, with a longest baseline of 97\,m. The wide-band correlator (WideX) was used, which provides a fixed spectral resolution of 1.95 MHz. For phase and amplitude calibration, the quasars 0400+258 and 0507+179 were used. For flux calibration, MWC349 was used for the three targets. The total time on source was 4.3 hr for J043903, 3.9 hr for J044148, and 3.6 hr for J043814. For all the scans, spurious signals were identified and flagged, and natural cleaning was performed to obtain the final image. The obtained resolution was around $\sim$3\arcsec.8$\times$4\arcsec.8 for all the targets. 

We do not resolve the disks and only J044148 was detected ($\gtrsim 3.5\,\sigma$). Figure~\ref{continuum_maps} shows the continuum maps at 3\,mm of the three targets. The total flux and its uncertainty were obtained from the image plane for each target. We also use {\tt uvmodelfit} in CASA to fit a point source, and find similar values as in the images for the total flux (Table~\ref{table_disks}).

We used recent ALMA observations\footnote{Project number \#2012.1.00743.S} in Band 7 (339 GHz, 0.89\,mm), for which the disks are detected with significant signal-to-noise,  to calculate values and limits of the spectral index. The sizes of the regions used to extract the fluxes was well within the maximum recoverable scale for all the ALMA observations. With these observations, we double the current number of spectral index estimates for BD disks. We took the 0.89\,mm measurements from \cite{ward2017}, where the details of the ALMA observations can be found.

\subsection{Millimeter Spectral Index and Disk Dust Mass}\label{sect:spectral_index}

\begin{figure}
 \centering
 \includegraphics[width=9.0cm]{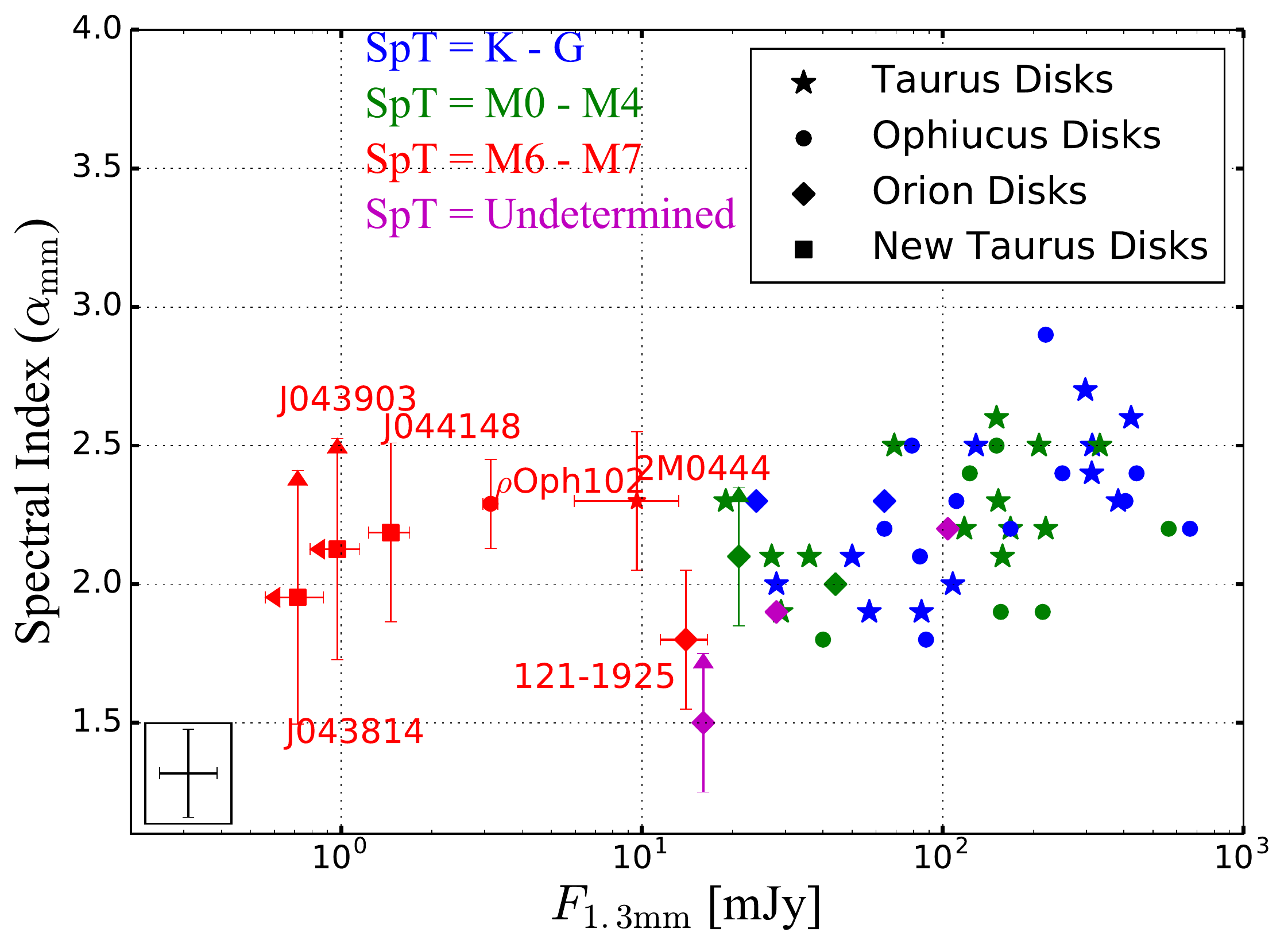}
  \caption{Spectral index of disks in different star-formation regions. The data for Taurus, Ophiucus, and Orion groups were taken from \cite{ricci2012b, ricci2013, ricci2014}. The new data reported are represented in squares. The error bar for most of the points is illustrated in the left corner, otherwise is plotted. }
   \label{spectral_index}
\end{figure}

With the total flux at 0.89 and 3.0\,mm, the integrated spectral index is given by $\alpha_{\rm{mm}} = \ln(F_{\nu_1}/F_{\nu_2}) /\ln (\nu_1/\nu_2)$, and the values for each of the BD disks are summarized in Table~\ref{table_disks}. To compare with observations of other protoplanetary disks in different star-formation regions, we gathered the data presented in \cite{ricci2012b, ricci2013, ricci2014} and classified it according to their spectral type. Figure~\ref{spectral_index} shows the spectral index (calculated either between $\sim$0.89 and 3 or $\sim$1.3 and 3\,mm) as a function of the flux at 1.3\,mm \citep[which is indicative of dust mass for optically thin emission, see e.g.][]{testi2014}. In the Rayleigh-Jeans regime of the spectrum, the calculation of the spectral index is independent of the used wavelengths. For the three observed BDs, we calculate the expected flux at 1.3\,mm with the obtained values of the spectral index, in order to have all fluxes at the same wavelength. The error bar of the flux in Fig.\ref{spectral_index} includes the uncertainties of the spectral index and the fluxes at 0.89 and 3\,mm.  The current data reveal low values of the spectral index ($\alpha_{\mathrm{mm}}\lesssim 3$), as observed for other BD disks and disks around more massive stars in different star-formation regions \cite[e.g.][]{ricci2012b, testi2014}. Nonetheless, two of the obtained values are lower limits (Table~\ref{table_disks}). 

%%%%%%%%%%%%
%FIGURE  3
%%%%%%%%%%%%
\begin{figure*}
 \centering
 \tabcolsep=0.05cm 
   \begin{tabular}{cc}   
   	\includegraphics[width=9cm]{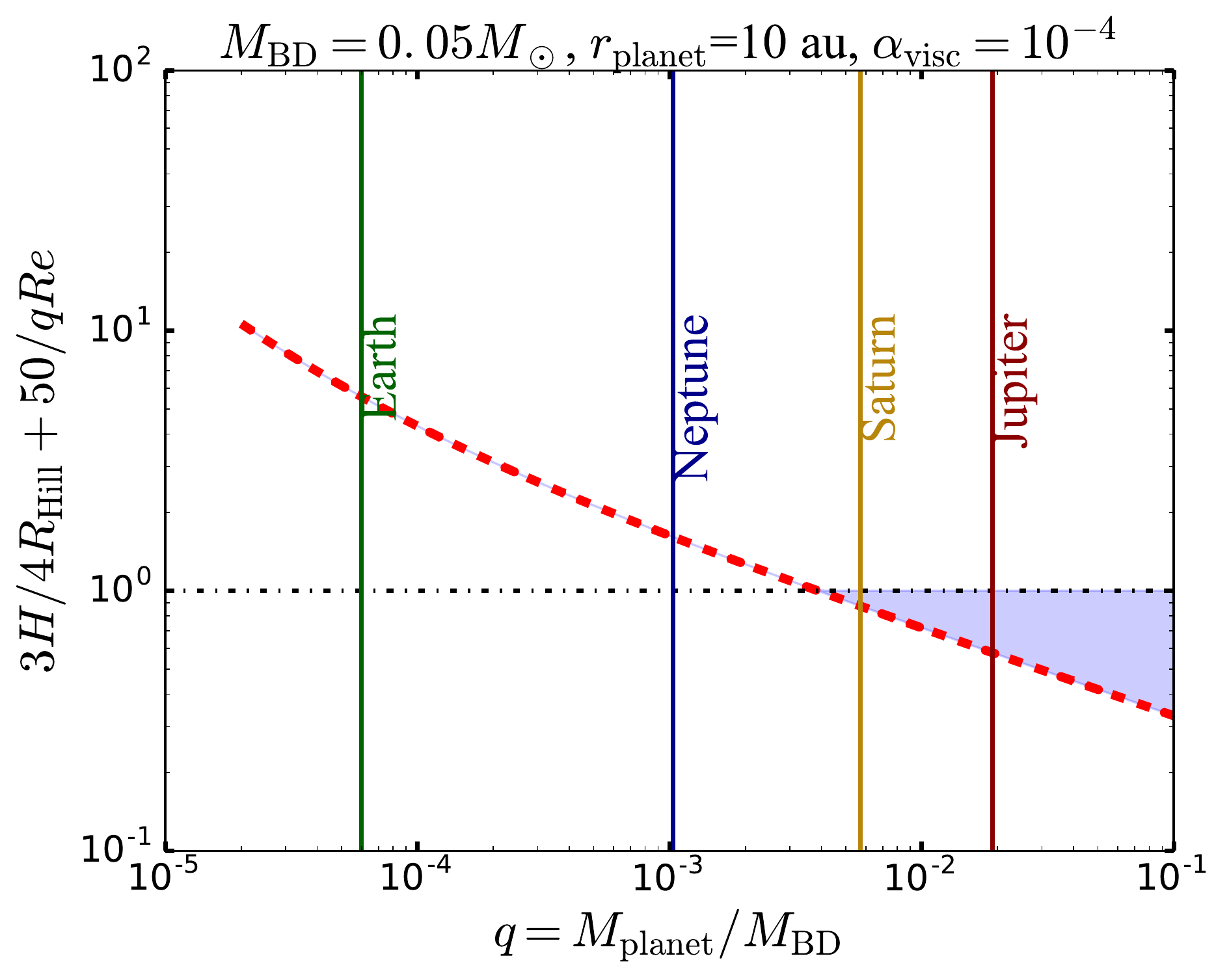}&
	\includegraphics[width=9cm]{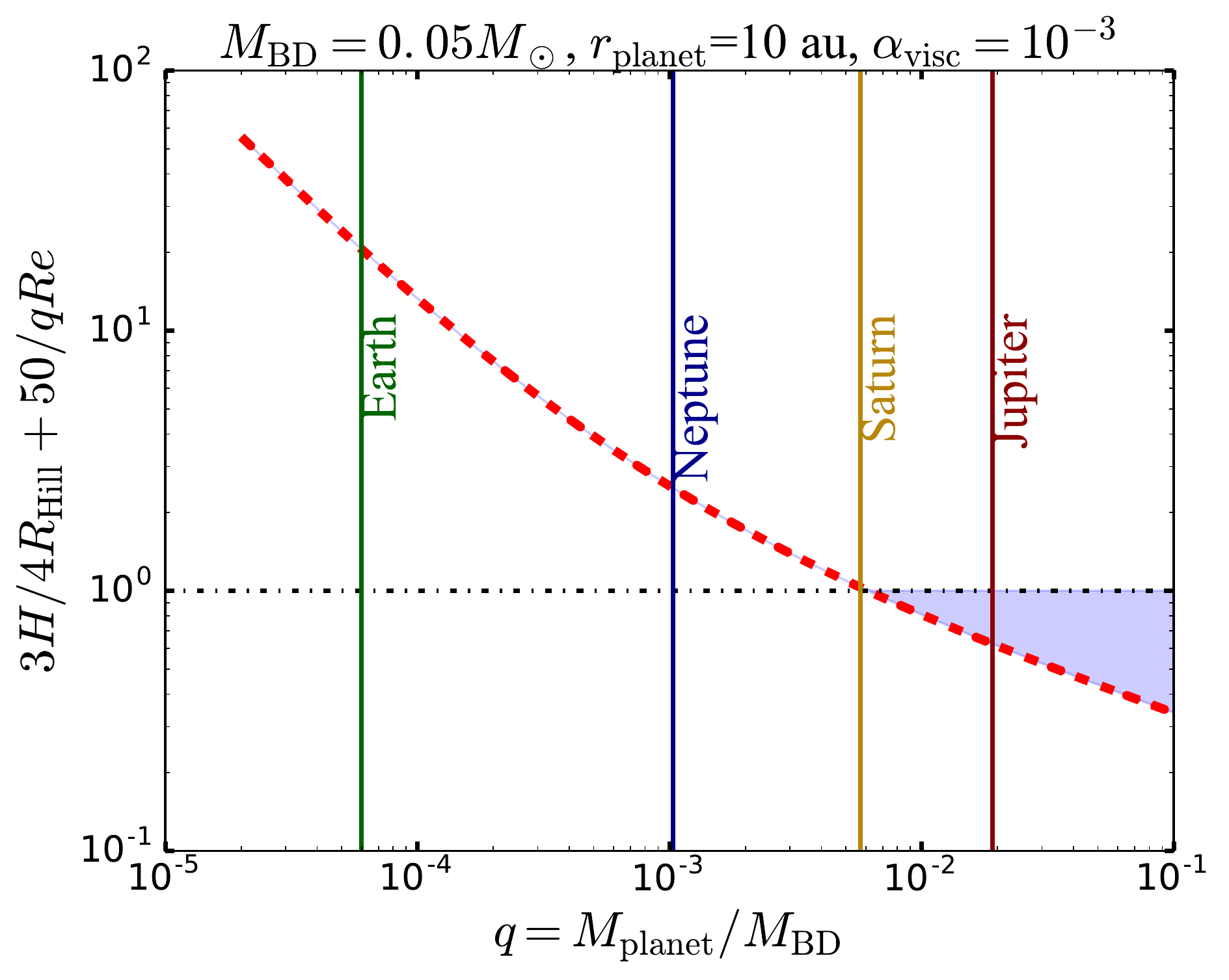}
   \end{tabular}
   \caption{Criterion for gap opening (Eq.~\ref{eq:crida}) assuming the parameters of BD disks (Table 2), two values of $\alpha_{\mathrm{visc}}$, and a planet position of 10\,au. In the shaded area the criterion is satisfied.}
   \label{crida_criterium}
\end{figure*}

The spectral index is indicative of grain size (with low values, $\alpha_{\mathrm{mm}}\lesssim 3$, implying millimeter grains in the outer parts of disks) as long as the emission is optically thin.  To estimate the validity of this assumption, we determine how compact the disk should be, if it hosts only large grains, for the millimeter emission to be optically thick. We obtain a very small outer disk radius ($\lesssim0.1\,$au), supporting our assumption of optically thin emission. 

With the 0.89\,mm fluxes, we estimate the disk dust mass assuming optically thin emission \citep{hildebrand1983, andrews2013}:

\begin{equation}
	M_{\mathrm{disk, dust}}\simeq\frac{{d^2 F_\nu}}{\kappa_\nu B_\nu (T(r))},
  \label{mm_dust_mass}
\end{equation}

\noindent where $d$ is the distance to the targets (taken to be 140\,pc), \ $\kappa_\nu$ is the mass absorption coefficient at a given frequency. We assume a frequency-dependent relation given by  $\kappa_\nu=2.3\,$cm$^{2}$\,g$^{-1}\times(\nu/230\,\rm{GHz})^{0.4}$ \citep[][]{andrews2013}. $B_\nu (T_{\rm{dust}})$ is the Planck function for a given dust temperature $T_{\rm{dust}}$, for which we assume the relation $T_{\rm{dust}}\approx25\times(L_\star/L_\odot)^{0.17}$\,K obtained by  \cite{vanderplas2016} for spectral types of M5 and later and a disk outer radius of 60\,au. The estimations for the disk dust mass for each target are shown in Table~\ref{table_disks}. 

The low values of the spectral index indicate that dust particles have millimeter sizes in these BD disks. In the next section, we investigate dust evolution models assuming a massive planet embedded in the outer disk to trap millimeter grains and compare the theoretical predictions of the spectral indices and millimeter fluxes with current observations.

\section{Dust Trapping By an Embedded Planet in a BD Disk} \label{sect:models}
From our current dust evolution models, millimeter grains around BDs can only be explained under extreme conditions, such as strong pressure inhomogeneities of around 40-60\% of amplitude \citep{pinilla2013}. However, such strong pressure bumps are not expected from magnetorotational instability (MRI) simulations, which predict pressure bumps with a maximum of 20-25\% of amplitude compared to the background density \citep[e.g.][]{uribe2011, dittrich2013, simon2014}. Strong pressure bumps can originate at the outer edge of a gap carved by a planet, and in this section, we aim to understand if trapping in BD disks due to an embedded planet can lead to low values of the spectral index as observed in BD disks. 

\paragraph{Gap opening criterion in BD disks.} The first question to investigate is, what is the minimum mass of a planet needed to open a gap in a disk around a BD? To answer this question, we use the gap opening criterion by \cite{crida2006}, which considers the disk  viscous torque, the gravitational torque from the planet, and the pressure torque. The criterion is

\begin{equation}
\frac{3}{4}\frac{H}{R_H}+\frac{50}{q Re}\lesssim 1
\label{eq:crida}
\end{equation}

\noindent where $Re$ is the  Reynolds number at the position of the planet $r_p$, which is equal to $r_p\Omega_p/\nu$ ($\Omega$ is the Keplerian frequency and $\Omega_p$ is calculated at the planet position), with $\nu$ being the disk viscosity, usually parametrized as $\nu=\alpha_{\rm{visc}}c_s^2/\Omega$ \citep{shakura1973}, and $c_s$ being the sound speed. In addition, $q$ is the planet-to-star mass ratio, $H$ is the disk aspect ratio equal to $c_s/\Omega$, and $R_H$ is the Hill radius of the planet, i.e. $r_H=r_p(q/3)^{1/3}$.

%%%%%%%%%%%%
%TABLE 2
%%%%%%%%%%%%
\begin{table} 
\label{table_parameters_BD}     
\centering                         
\begin{tabular}{cc}       
\hline\hline                 
Parameter & Values \\    
\hline 
   $M_{\rm{BD}} [M_\odot]$ & $0.05$  \\
   $L_{\rm{BD}} [L_\odot]$ & $0.03$  \\  
   $T_{\rm{BD}}$ [K]&$2880$\\ 
   $M_{\mathrm{disk}} [M_{\mathrm{Jup}}]$ & $2.0$  \\      
   $R_{\mathrm{in}}$[AU] &2.0\\                   
   $R_{\mathrm{out}}$[AU] &$60$\\
   $\alpha_{\mathrm{visc}}$ & $10^{-3}$  \\  
   $v_f$ [m\,s$^{-1}$]& $30$\\
\hline
\hline                     
\end{tabular}  
\caption{Assumed parameters for the central BD and the disk. For the BD, we took the values of the BD $\rho$-Oph 102 \citep{ricci2012}, which are similar to the values of the BD disks observed with PdBI and reported in Table.~\ref{table_disks}}
\end{table}

Because the BD disks are colder than T Tauri disks and the mass of the central star is also lower, the required planet mass for a planet to open a gap differs from the one needed around a T Tauri star (typically with a $q$ value of $10^{-3}$, which corresponds to a 1\,$M_{\rm{Jup}}$ around a Sun-like star). We calculate the opening gap criterion (Eq.~\ref{eq:crida}), assuming the BD parameters  as in Table~2, two values of $\alpha_{\rm{visc}}$ ($\alpha_{\rm{visc}}=[10^{-4}, 10^{-3}]$) \citep[in agreement with the level of turbulence inferred from observations and MRI simulations;][]{fromang2006, flaherty2015, teague2016}, and a simple temperature parametrization. As proposed by \cite{kenyon1987}, we assume a power law that depends on the  temperature ($T_{\star}$) and radius ($R_{\star}$) of the central star, or in our case of the BD, given by

%%%%%%%%%%%%
%FIGURE 4
%%%%%%%%%%%%
\begin{figure*}
 \centering
 \tabcolsep=0.05cm 
   \begin{tabular}{cc}   
   	\includegraphics[width=9.5cm]{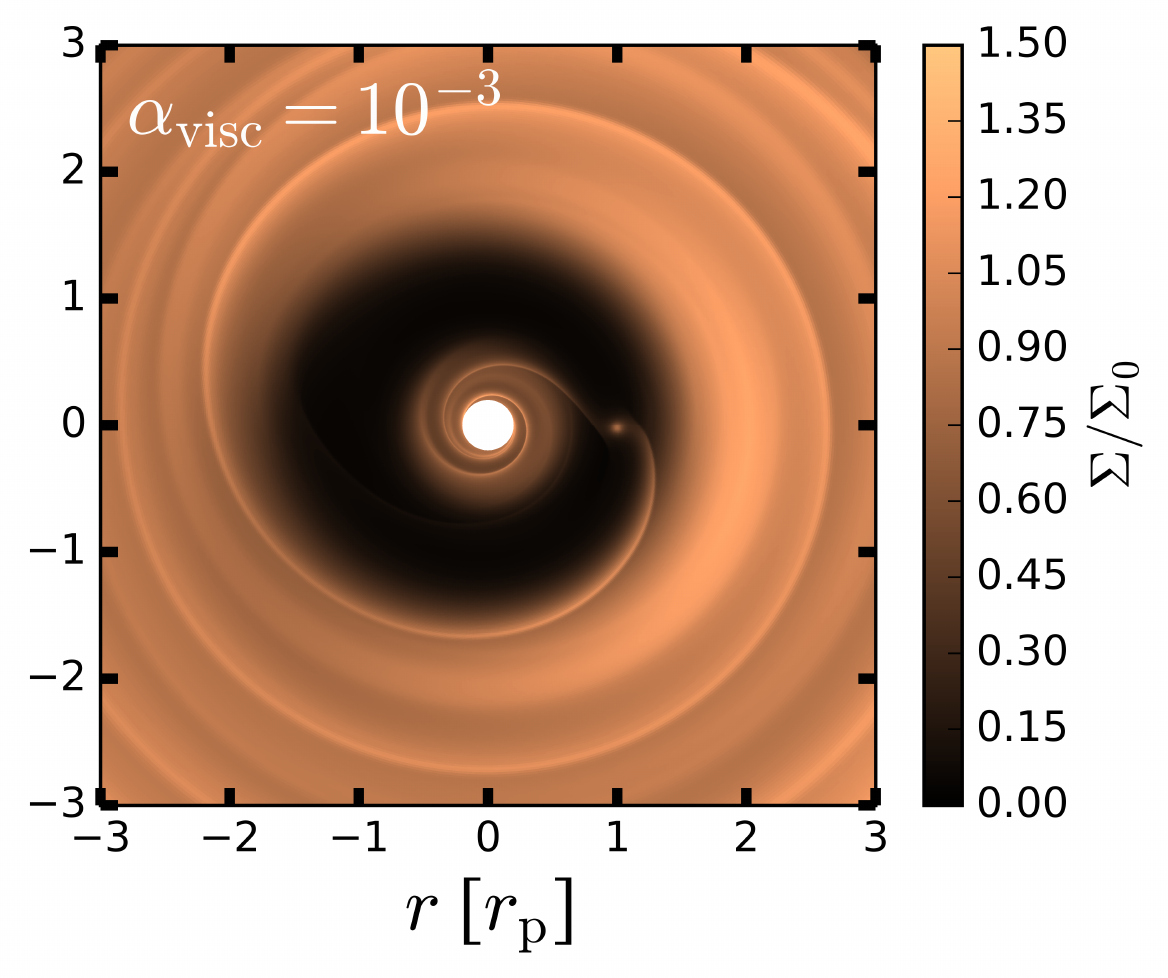}&
	\includegraphics[width=8.5cm]{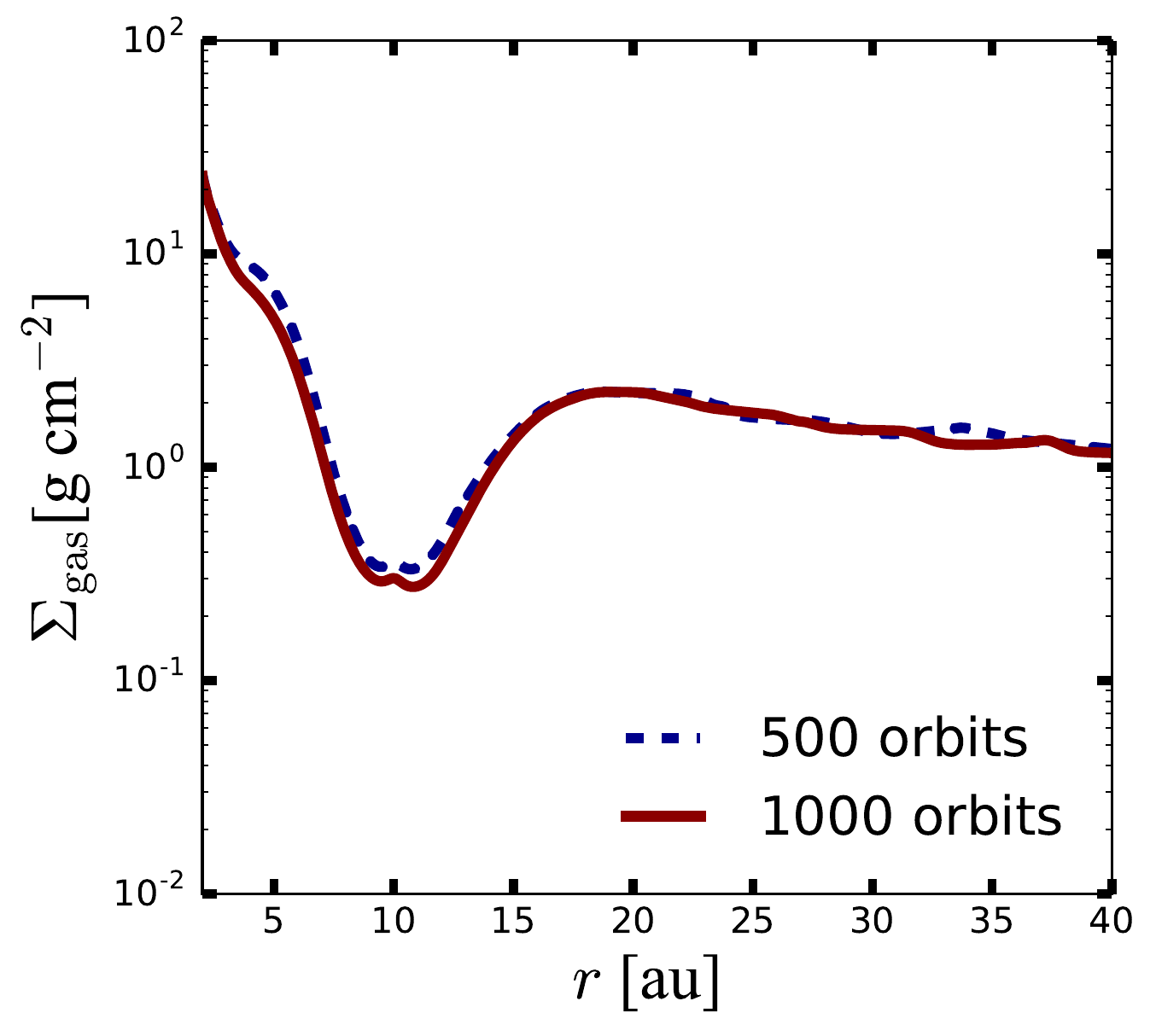}	 	
   \end{tabular}
   \caption{Left panel: 2D gas density distribution after 1000 orbits of evolution assuming a single planet and the parameters as in Table~2. The gas surface density is normalized to the initial profile. Right panel: the averaged gas surface density after 500 and 1000 orbits, which is later assumed to compute the dust evolution.}
   \label{hydro_simulations}
\end{figure*}

\begin{equation}
	T(r)=T_{\star}\left(\frac{R_{\star}}{r}\right)^{1/2} \phi_{\mathrm{inc}}^{1/4},
  	\label{eq:temperature} 
\end{equation}

\noindent  where the angle between the incident radiation and the local disk surface is taken to be $\phi_{\mathrm{inc}}=0.05$, to assume temperatures close to the disk midplane \citep[e.g.][]{hartmann1998}.

The criterion to open a gap (Eq.~\ref{eq:crida}) is satisfied for a Saturn-like planet located at $r_p=10$ au and for $\alpha_{\mathrm{visc}}=10^{-3}$ (Fig.~\ref{crida_criterium}). This corresponds to a planet to star mass ratio of $q=6\times10^{-3}$. Note that this mass ratio is lower than the one observed in 2MASSWJ\,1207334-393254 \citep{chauvin2004}. The formation mechanisms of this planetary mass companion around the BD was explored by \cite{lodato2005}, concluding that the core accretion is too slow to form this planet. However, alternative mechanisms, such as gravitational instability and binary formation are challenging, but still possible. For a lower viscosity ($\alpha_{\mathrm{visc}}=10^{-4}$), a lower-mass planet can open a gap (0.7\,$M_{\rm{Saturn}}$), because the viscous torque is lower. At a more distant location, the planet would be more massive to satisfy the condition. These proposed planets cannot be detected with our current observational capabilities, neither by means of radial velocity or direct imaging. \cite{pinilla2014}  analyzed the potential correlation between the planet position (which is directly connected to the cavity size in transition disks) and the spectral index, finding a positive correlation (observationally and theoretically), given by $\alpha_{\rm{mm}}=0.011\times R_{\rm{cav}}+2.36$, which implies that the spectral index would not change significantly if the planet is located on a different orbit. This dependence can be explained by the fact that moving the planet inwards increases the maximum size of particles inside the trap,  but also reduces the surface area of the dust trap.

\paragraph{Hydrodynamical simulation.}We assume a planet of Saturn mass located at 10\,au to run hydrodynamical simulations, in order to study the planet-disk interaction process in a locally isothermal BD disk, assuming a value of $\alpha_{\mathrm{visc}}=10^{-3}$ (Table~2). The motivation to select 1\,$M_{\rm{Saturn}}$ for the planet mass is because the masses of the disks around BD are of the order of few Jupiter masses, and if the planets are formed within the disk, we thus select the minimum mass planet to open a gap and to have effective trapping of particles. We do not consider the case of $\alpha_{\mathrm{visc}}=10^{-4}$, because when turbulent diffusion of grains is too low, fragmentation of particles is unlikely inside a pressure trap, leading to the formation of planetesimals in million-year timescales and the depletion of millimeter-sized pebbles. As a consequence, dust evolution models assuming low viscosity, usually lead to large planetesimals with low opacity, thus low millimeter fluxes and values of the spectral indices that are in disagreement with observations \citep[e.g.][]{pinilla2013, ovelar2016}.

%%%%%%%%%%%%
%FIGURE 5
%%%%%%%%%%%%
\begin{figure*}
 \centering
   \begin{tabular}{cc}   
   	\includegraphics[width=8.4cm]{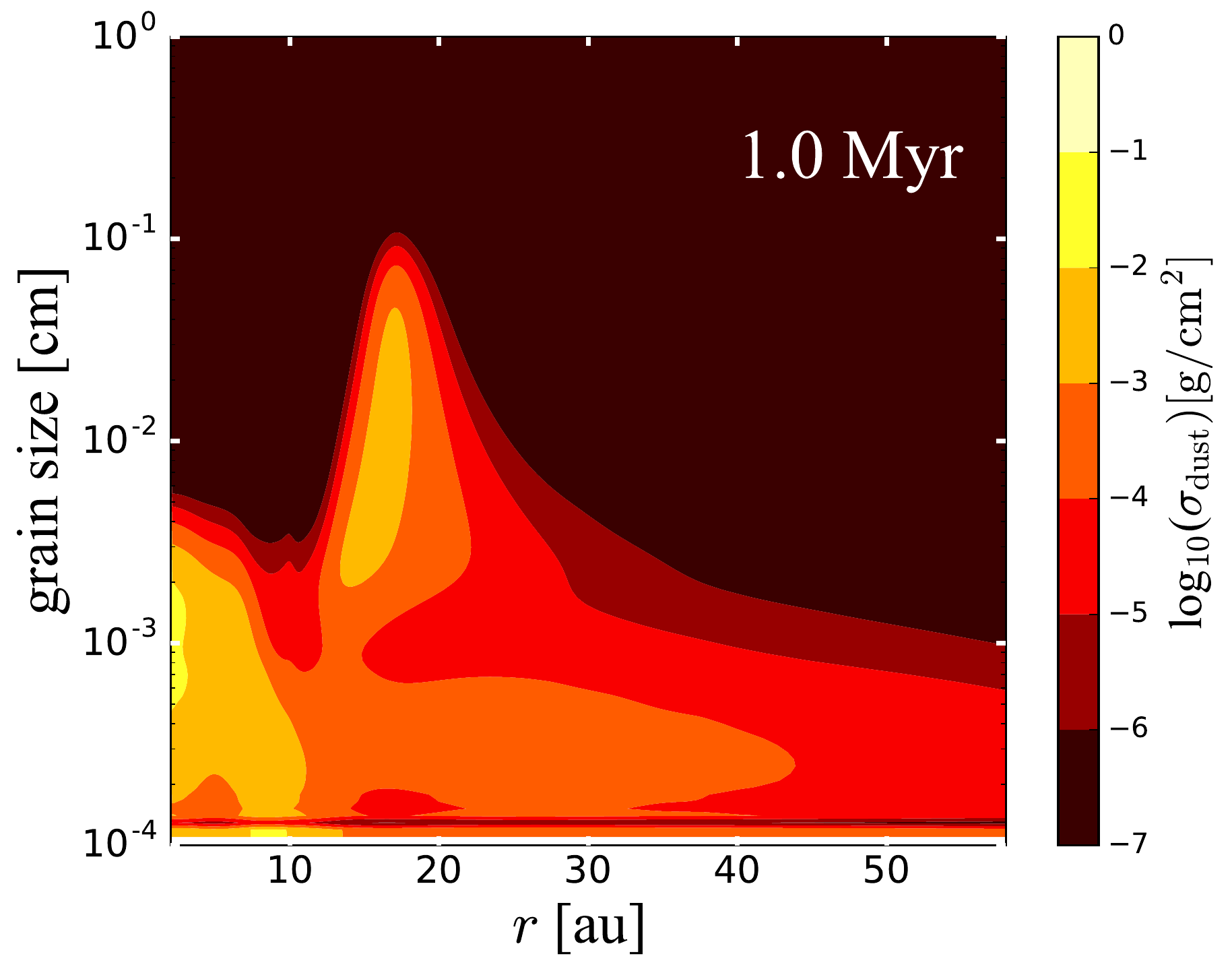}&
	\includegraphics[width=8.6cm]{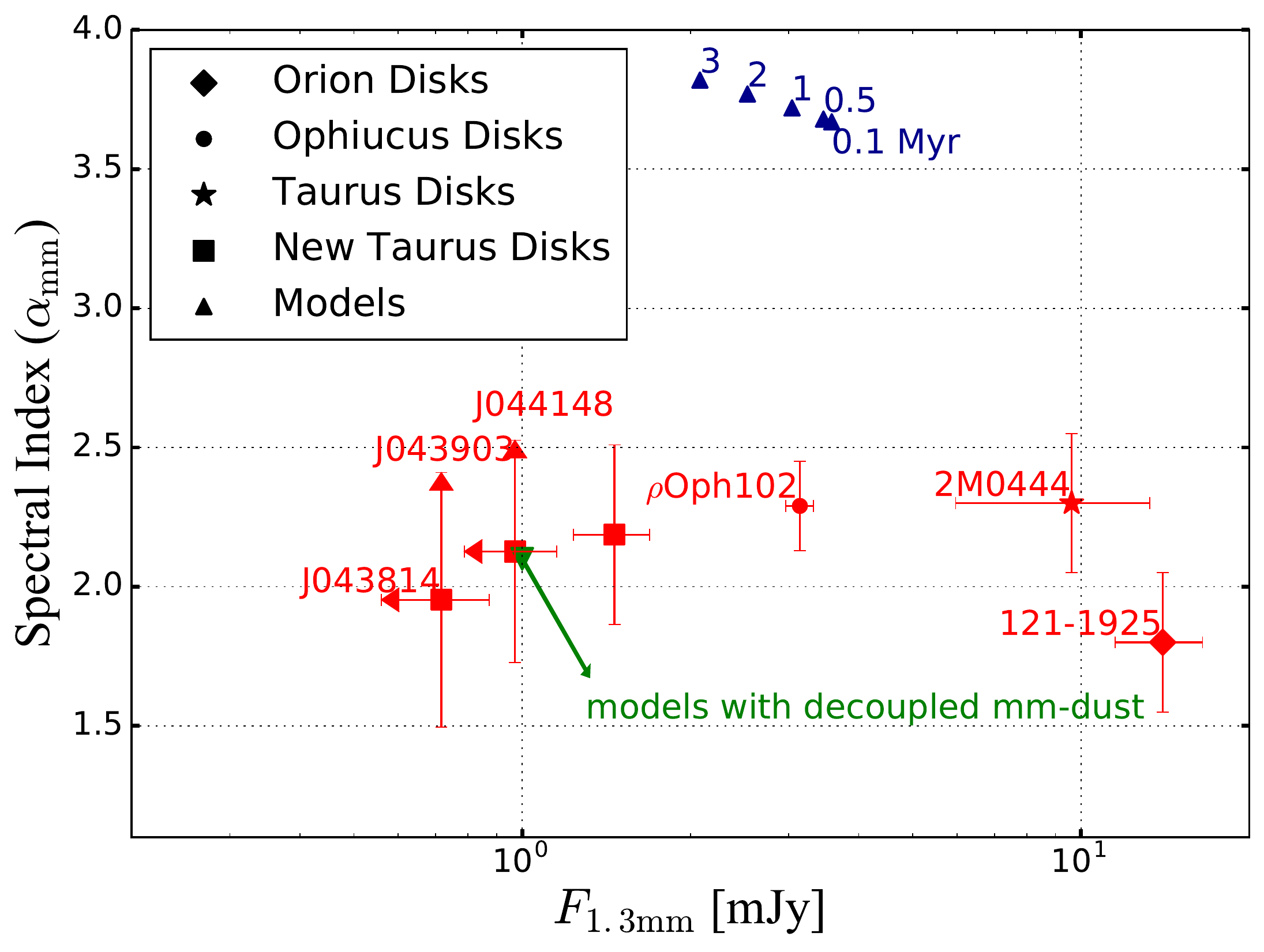}
   \end{tabular}
   \caption{Left panel: vertically integrated dust density distribution ($\sigma_{\rm{dust}}$) as a function of the grain size and radius, assuming the gas surface density profile from Fig.~\ref{hydro_simulations} after 1\,Myr of evolution. Right panel: spectral indices obtained from millimeter observations of BD disks compared to models of dust trapping by an embedded massive planet (blue triangles). The green triangle corresponds to a test where the gas density is low and thus millimeter-sized particles are decoupled from the gas (Fig~\ref{dust_simulations_low_mass_disk}).}
   \label{dust_simulations}
\end{figure*}

For the hydrodynamical models, we used the legacy version of FARGO2D \citep{masset2000}. For the initial gas surface density, we assumed a power law, such that $\Sigma\propto r^{-1}$. In \cite{pinilla2013}, we demonstrated that changing the power-law index does not have a significant effect on the final dust density distributions and the resulting spectral indices.  In addition, we assume a logarithmically spaced radial grid with 512 points from 0.2 to 6\,$r_p$. The azimuthal grid is linear with 950 grid cells. The scale height and the disk temperature are assumed consistently with Eq.~\ref{eq:temperature}, that is, the disk aspect ratio is increasing with radius as $H/r=h_0\times (r/r_p)^f$, with a flaring index $f=0.25$. The disk aspect ratio at the planet position ($h_0$) is taken to be 0.14. The mass of the disk is taken to be $2\,M_{\mathrm{Jup}}$. This disk mass value is an optimistic value compared to the values reported in Table~\ref{table_disks} (and assuming a dust-to-gas disk mass ratio of 1/100), and we assume a disk mass of a BD disk similar to $\rho$\,Oph102 and 2M0444 \citep{ricci2013, ricci2014}. Decreasing the disk mass reduces the maximum grain size and hence it increases the resulting spectral index \citep{birnstiel2010a}.

Results from FARGO simulations are presented  in Fig.~\ref{hydro_simulations}, which shows the 2D gas density distribution after 1000 local orbits ($\sim3\times10^{4}$ years) and the averaged gas surface density after 500 and 1000 orbits. The resulting gap is shallower than in the case of a T Tauri star. For instance, in the case of $q=6\times10^{-3}$, the gas surface density inside the gap for a T Tauri or Herbig disk, with the same $\alpha_{\rm{visc}}$, is more depleted than in the BD disk \citep[see, e.g.][where the gas surface density is depleted by 4 orders of magnitude for the same $\alpha_{\rm{visc}}$ and $q$]{pinilla2015}. The main reason for the shallower gap in the case of BD disks is due to the higher scale height at the planet position. The scale height depends on the sound speed ($c_s$) and Keplerian frequency ($\Omega$), and because of  the lower central mass, the scale height at the planet position is closer to the planet Hill's radius ($r_H$), making the equilibrium shape of the gap shallower \citep{crida2006}. Our simulations show that the gas surface density inside the gap decreases by around one order of magnitude. There are some observational suggestions that disks around BDs are actually flatter \citep[e.g.][]{pascucci2009, szucs2010,liu2015, daemgen2016}, but this depends on the range of wavelengths that is used as an indicator for the degree of settling in the disk \citep{furlan2011, mulders2012}. However, observations provide information about the dust settling and not directly about the gas disk scale height.

\paragraph{Dust evolution and theoretical spectral indices.} We used the azimuthally averaged gas surface density from the hydrodynamical simulation after the disk has reached a steady-state ($\sim$1000 orbits) to self-consistently calculate the dust density distribution, taking into account dust dynamics and the dust growth process \citep[coagulation, fragmentation, and erosion, see][]{birnstiel2010}. This combination of hydrodynamical and dust evolution models was introduced in \cite{pinilla2012}, and we refer to this paper for more details. We follow the evolution of 180 grain sizes from 1\,$\mu$m to 2\,m. Initially, all the dust is taken to be 1\, $\mu$m sized particles distributed  in the disk with a constant gas-to-dust ratio of 100. We assume a fragmentation velocity of particles of 30\,m\,s$^{-1}$, and also that the collision at this velocity or higher will lead to fragmentation of particles. These values are in agreement with numerical and laboratory experiments of dust particles with water ice mantles  \citep[e.g.][]{paszun2006, wada2009, wada2011, gundlach2015}, consistent with the disk temperatures that we assumed (from 55 to 10\,K from the inner to the outer radius). We model the evolution from $3\times10^{4}$ years up to 1\,Myr.

The results of the dust density distribution as a function of radius and grain size are shown in the left panel of Fig.~\ref{dust_simulations}. The dust distribution shows an accumulation of particles at the location of the gas surface density  maximum (or pressure maximum), at $\sim17\,$au. In this case, the maximum grain size in the whole disk is determined by radial drift because particles drift inwards before they can grow to larger sizes. As consequence, it is only inside a pressure bump that particles can grow to millimeter sizes in BD disks \citep[e.g.][]{pinilla2013, birnstiel2015}. Inside the pressure bump, the particles reach sizes of $\sim1\,$mm. Outside the pressure bump, the maximum grain size is smaller ($\sim 0.08\,$mm), because the high radial drift also leads to the fragmentation of particles. 

%%%%%%%%%%%%
%FIGURE 6
%%%%%%%%%%%%
\begin{figure*}
 \centering   
   	\includegraphics[width=18cm]{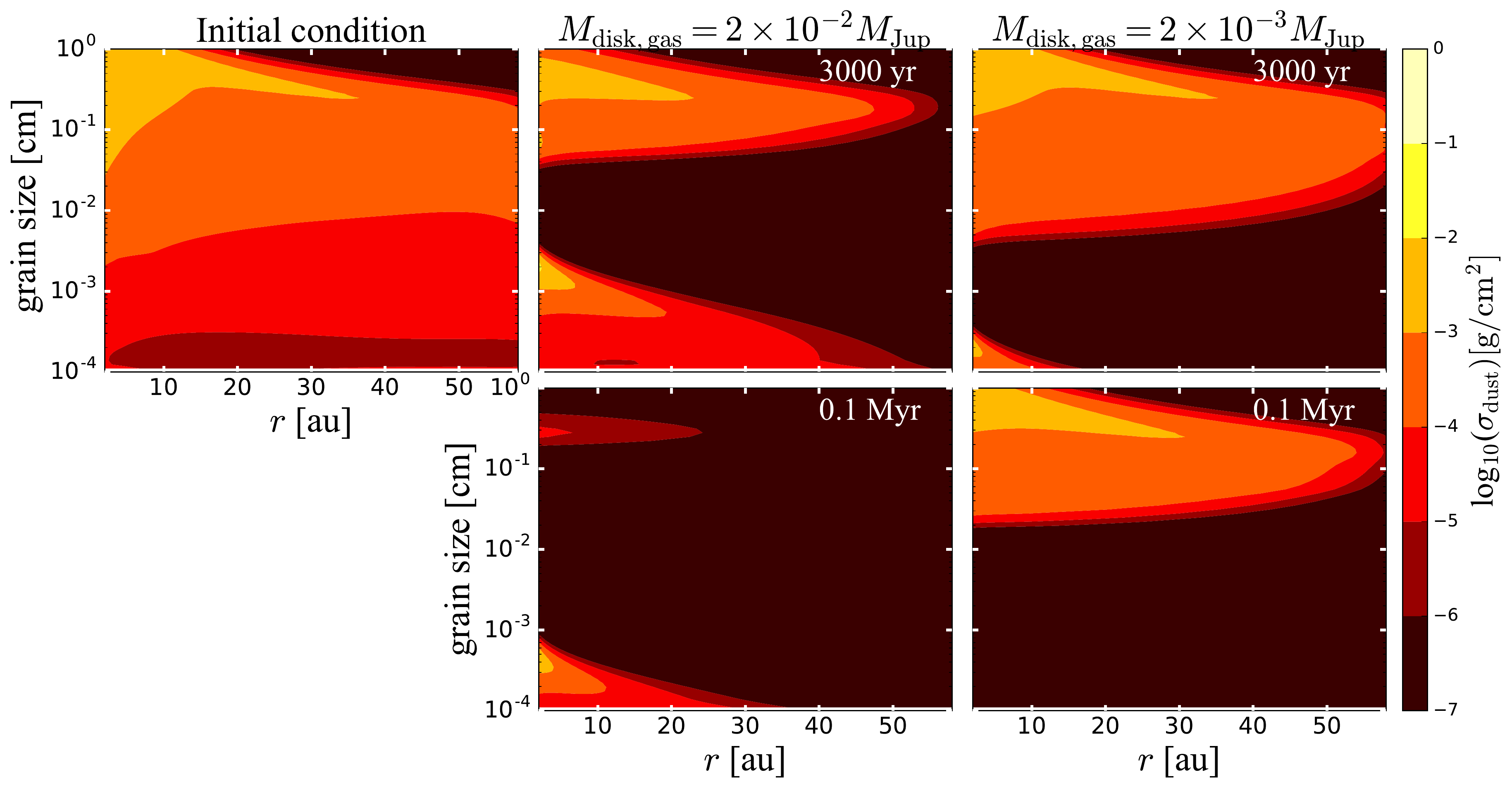}
   \caption{Left panel: initial dust density distribution as expected from grain growth without experiencing radial drift \citep{pinilla2013} and with the same disk conditions as in Table~2. Middle panels: dust density distribution after 3000\,years (top) and 0.1\,Myr (bottom) of evolution considering dust dynamics (drift, drag, and turbulent diffusion), but without dust coagulation processes and reducing the gas disk mass by a factor of 100. Right panels: same as middle panels but reducing the gas disk mass by a factor of 1000.}
   \label{dust_simulations_low_mass_disk}
\end{figure*}

In order to compare the millimeter fluxes that result from this dust evolution model, we take the dust density distribution at different times of evolution (0.1, 0.5, 1, 2, and 3\,Myr) and calculate the optical depth ($\tau_\nu$) at two different millimeter wavelengths, such that $\tau_\nu=\sigma(r,a) \kappa_\nu/\cos i$, where $\sigma(r,a)$ is the vertically integrated dust density distribution at a given time of evolution (as shown in Fig.~\ref{dust_simulations} at 1\,Myr). The opacity for each grain size and at a given frequency or wavelength ($\kappa_\nu$) is calculated following Mie theory, and we assume the volume fractions and optical properties for the dust particles as in \cite{ricci2010}; and $i$ is the disk inclination, which we assume to be $20^\circ$ \citep[as in][for the targets observed with PdBI]{scholz2006}. With the optical depth, we obtain the total flux as

\begin{equation}
	F_\nu=\frac{2\pi\cos{i}}{d^2}\int_{R_{\mathrm{in}}}^{R_{\mathrm{out}}} B_\nu(T(r)) [1-e^{-\tau_\nu}] r dr,
  \label{eq:flux}
\end{equation}

\noindent where $d$ is the distance to the source, which is assumed to be 140\,pc as the Taurus star-forming region, and $B_\nu(T(r))$ is the Planck function, for which we assume the temperature profile of Eq.~\ref{eq:temperature}. After we obtain the millimeter fluxes, we calculate the spatially integrated spectral index as explained in Sect~\ref{sect:spectral_index}, between 1.3 and 3.0\,mm. The results are shown in the right panel of Fig.~\ref{dust_simulations}, together with the available data of BD disks (see Fig.~\ref{spectral_index}).  

We find that the theoretical spectral indices are high at any time of evolution ($\gtrsim 3.5$) for the case of trapping by an embedded planet, in disagreement with current millimeter observations of BD disks. The spectral index only has values slightly lower than 3 at the location of the pressure maximum ($\sim17$\,au), but in a very narrow region (width of $\sim5$\,au). The dust concentration in the pressure maximum is very narrow because of the high radial drift that millimeter-sized particles experience in disks around BD. Contrary to the case of a disk around T Tauri or Herbig stars, there are too few millimeter-sized particles concentrated in a very narrow region when there is trapping by a planet in a BD disk. This leads to high values of the spectral index integrated over the entire disk at any time of evolution. The resulting millimeter fluxes are, however, in the range of some of the observed BD disks. Moving the planet or changing the initial disk parameters (gas surface density slope, fragmentation velocity, outer radius) would not contribute significantly to decreasing the spectral index based on the results from \cite{pinilla2013, pinilla2014}. As an experiment, we also assumed a BD disk for which the outer radius is at 25\,au (right beyond the bump of millimeter grains) to run the dust evolution and test if in this case the spectral index decreases. However, the spectral index barely changes compared to the model shown in Fig.~\ref{dust_simulations}, with $R_{\rm{out}}=60\,$au.

\section{Discussion} \label{sect:discussion}
Interferometric millimeter observations with CARMA, ALMA, and PdBI of disks around BDs provided the first measurements of the millimeter spectral index, and revealed values expected when dust grains have grown to millimeter sizes \cite[e.g.][]{ricci2013, ricci2014}. Such observations are a challenge for models of dust evolution in disks because the radial inward drift velocities are expected to be higher for particles in BD disks than in disks around T Tauri stars, depleting the disk in grains before they can grow \citep{pinilla2013}.

In models of dust evolution around T Tauri and/or Herbig Ae/Be stars, the problem of dust radial drift can be solved by assuming either several, and smooth, pressure bumps globally distributed in the disk, or with a single and strong pressure bump, as the one formed when a massive companion ($\gtrsim1\,M_{\rm{Jup}}$) is embedded in the disk, which can be the case for transition disks \citep[e.g.][]{pinilla2012, pinilla2014, espaillat2014, owen2016}. In the case of a BD disk, we demonstrated that a Saturn mass planet is the minimum mass required to open a gap in the gas surface density and trap millimeter-sized particles.

However, for any of the targets in Table~\ref{table_disks}, the disk mass is too low to form a Saturn-like planet at any location of the disk. By assuming a disk accretion rate of $10^{-12}$-$10^{-9}\,M_{\rm{\odot}}$\,yr$^{-1}$ \citep[values observed for BD disks,][]{herczeg2009, rigliaco2011}, a disk age of 1\,Myr, a dust-to-gas disk mass ratio of 1/100, and the disk dust masses obtained from the 0.89\,mm data for the BD disks in Table~\ref{table_disks}, we estimate that the initial disk mass is $\sim3-6\,M_{\rm{Saturn}}$ for the most massive disk in Table~\ref{table_disks} (J044148). Therefore, these BD disks do not have enough mass to form giant planets, as was shown by \cite{payne2007}. Nonetheless, some observations of proto-BD, suggest values of the envelop mass of $\sim20-30\,M_{\rm{Jup}}$ \citep{andre2012, palau2014}, which means that there may be enough material at earlier states to form a Saturn-like planet around a BD.

In addition, we show that the hypothesis of dust trapping by a massive planet is not suitable to explain the low values of the spectral indices of BD disks  ($\alpha_{\rm{mm}}\lesssim3.0$, Fig.~\ref{spectral_index}), contrary to the case of  more massive and warmer (transition) disks. Therefore, multiple and strong bumps remain as a favorable scenario to explain the current millimeter observations of BD disks \citep{pinilla2013}. Understanding what can be the origin of such strong pressure bumps in BD disks is still an open question, since multiple massive planets are unlikely. 

One possible solution to allow millimeter-sized particles to remain in BD disks is that their disk gas masses are much lower than assumed, such that millimeter grains do not experience radial drift. In this scenario, the millimeter grains may be completely decoupled from the gas and do not feel the headwind that arises from the sub-Keplerian gas velocity. As a numerical experiment, we assume a smooth dust density distribution as expected from grain growth without experiencing radial drift \citep{pinilla2013}, but with the same disk conditions as in Table~2, and reducing the gas surface density by a factor of 100 or 1000 (Fig.~\ref{dust_simulations_low_mass_disk}) and we do not consider the feedback from the dust to the gas (nor changes of the radial drift). This simple experiment aims to test if the millimeter-sized particles can remain in the outer disk and if the amount of dust is enough to have values of the spectral index and millimeter fluxes in agreement with observations. 

In these tests, the coupling parameter of dust particles with the gas for the millimeter grains increases by two or three orders of magnitude, such that they are partially or completely decoupled. With the assumed dust density distribution, we model the dust evolution considering only the dust dynamics (drift, drag, and turbulent diffusion), and neglecting coagulation processes (Fig.~\ref{dust_simulations_low_mass_disk}). However, including fragmentation does not change the results because the millimeter-sized particles are beyond the fragmentation barrier. 

In the case where the gas surface density is reduced by a factor of 100 and the disk gas mass is $2\times10^{-2}\,M_{\rm{Jup}}$, the millimeter-sized particles only remain in the disk for short times of evolution. This is a result of dust diffusion, which can still affect the millimeter-sized particles that are partially decoupled, moving the grains to the inner regions where drift can be more effective. However, when the gas surface density is reduced by a factor of 1000 and the disk gas mass is $2\times10^{-3}\,M_{\rm{Jup}}$, the millimeter-sized particles are completely decoupled from the gas and they are not affected by diffusion as in the previous case. As a consequence, these particles can remain in the entire disk for long times of evolution while small grains ($\lesssim0.1$\,mm) drift inwards. 

Considering the dust density distribution after 0.1\,Myr, we calculate the millimeter fluxes and spectral indices as explained in Sect.~\ref{sect:spectral_index}. We obtain a flux at 1.3\,mm of $\sim1$\,mJy and a spectral index of $\alpha_{\rm{mm}}=2.1$ in agreement with the observed values (Fig.~\ref{dust_simulations}). The crucial question to address with future observations is to determine the gas mass in disks around BDs in order to test this idea. 

An alternative possibility is fluffy dust growth, which avoids barriers of planetesimal formation, such as radial drift and fragmentation \citep{kataoka2013}. In this scenario, fluffy aggregates grow with a low filling factor such that large aggregates (meter to kilometer-sized particles) can still be coupled to the gas and avoid the radial drift. They are afterwards compressed by collisional compression,  self-gravity, and disk gas to form more compact planetesimals. The opacity of fluffy aggregates is  expected to be similar to compact dust grains at millimeter-wavelengths. In particular, it is quite difficult to distinguish between both scenarios from spatially-integrated values of the spectral index \citep{kataoka2014}. Spatially resolved observations with high angular resolution that radially resolve the spectral index is an alternative to discern between compact and fluffy growth and a challenge for future observations, in particular in the context of low-mass, cold, and compact disks around BDs.

\section{Conclusions} \label{sect:conclusion}
We present 3\,mm continuum observations carried out with the PdBI of three BD disks in the Taurus star forming region, which have been observed with ALMA in Band 7 ($0.89$\,mm). Based on these data, we obtained the values and lower limits of the spectral index, finding low values ($\alpha_{\rm{mm}}\lesssim3.0$), which suggest the presence of millimeter-sized particles in these BD disks as observed for disks around T Tauri and Herbig Ae/Be stars in different star-formation regions. In addition, we calculate the dust disk masses, finding values between 0.4 and 2.3\,$M_{\oplus}$.

We compare these observations with models of dust evolution when a massive planet is embedded in a disk around a BD, in order to study particle trapping and the expected spectral indices. We found that a Saturn mass planet is the minimum mass required to open a gap in the gas surface density and to trap millimeter-sized particles. This scenario is,  however, an unlikely explanation because Saturn mass planets are far above the present-day planet forming capabilities of these disks. One possibility is that disks were more massive and these planets formed early on or these companions form as a binary systems. In addition, we obtained the theoretical spectral indices from dust evolution models assuming a Saturn mass planet embedded in a BD disk. We found, however, that the spectral indices are high ($\gtrsim3.5$), in disagreement with current millimeter observations of BD disks.

An alternative possibility to current observations is that the gas mass in BD disks is so low ($\sim$2$\times10^{-3}\,M_{\rm{Jup}}$), such that the millimeter-sized particles are completely decoupled, and preventing them from drifting inward. However, there are indications that some BD disks are gas rich \citep{ricci2014}, but a more systematic and sensitive survey with ALMA that provide information of gas distribution/mass, is required to solve this question.

A plausible explanation to current observations is strong pressure bumps distributed in the entire disk and not only a local strong pressure bump \citep{pinilla2013}. Therefore, it is possible that the low spectral indices observed in BD disks hint to unresolved multi-ring substructures, as observed in more massive and warmer disks \citep[e.g. HL\,Tau,][]{alma2015}.  High angular resolution and sensitive observations with ALMA that allow us to detect dust rings in BD disks, are crucial to understand the drift barrier in the extreme conditions of BD disks. \\

\acknowledgements{We thank D.~Apai, T.~Bergin, J.~Williams, and A.~Youdin for interesting discussions about this project. P. P. acknowledges support by NASA through Hubble Fellowship grant HST-HF2-51380.001-A awarded by the Space Telescope Science Institute, which is operated by the Association of Universities for Research in Astronomy, Inc., for NASA, under contract NAS 5- 26555. M.B. and  G.v.d.P.  acknowledge funding from ANR of France under contract number ANR-16-CE31-0013 (Planet Forming Disks). This paper makes use of the following ALMA data: ADS/JAO. ALMA \#2012.1.00743.S. ALMA is a partnership of ESO (representing its member states), NSF (USA) and NINS (Japan), together with NRC (Canada),  NSC and ASIAA (Taiwan), and KASI (Republic of Korea), in cooperation with the Republic of Chile. The Joint ALMA Observatory is operated by ESO, AUI/NRAO, and NAOJ.}


\begin{thebibliography}{}
\bibitem[Alexander \& Armitage(2007)]{alexander2007} Alexander, R.~D., \& Armitage, P.~J.\ 2007, \mnras, 375, 500 

\bibitem[ALMA Partnership et al.(2015)]{alma2015} ALMA Partnership, Brogan, C.~L., P{\'e}rez, L.~M., et al.\ 2015, \apjl, 808, L3 

\bibitem[Andr{\'e} et al.(2012)]{andre2012} Andr{\'e}, P., Ward-Thompson, D., \& Greaves, J.\ 2012, Science, 337, 69 


\bibitem[Andrews et al.(2013)]{andrews2013} Andrews, S.~M., Rosenfeld, K.~A., Kraus, A.~L., \& Wilner, D.~J.\ 2013, \apj, 771, 129 

\bibitem[Andrews et al.(2016)]{andrews2016} Andrews, S.~M., Wilner, D.~J., Zhu, Z., et al.\ 2016, \apjl, 820, L40 

\bibitem[Apai et al.(2005)]{apai2005} Apai, D., Pascucci, I., Bouwman, J., et al.\ 2005, Science, 310, 834 

\bibitem[Bate(2009)]{bate2009} Bate, M.~R.\ 2009, \mnras, 392, 590 

\bibitem[Bate(2012)]{bate2012} Bate, M.~R.\ 2012, \mnras, 419, 3115 

\bibitem[Bayo et al.(2017)]{bayo2017} Bayo, A., Joergens, V., Liu, Y., et al.\ 2017, \apjl, 841, L11 


\bibitem[Beckwith et al.(2000)]{beckwith2000} Beckwith, S.~V.~W., Henning, T., \& Nakagawa, Y.\ 2000, Protostars and Planets IV, 533 

\bibitem[Birnstiel et al.(2010a)]{birnstiel2010a} Birnstiel, T., Ricci, L., Trotta, F., et al.\ 2010a, \aap, 516, L14 


\bibitem[Birnstiel et al.(2010b)]{birnstiel2010} Birnstiel, T., Dullemond, C.~P., \& Brauer, F.\ 2010b, \aap, 513, A79 

\bibitem[Birnstiel et al.(2015)]{birnstiel2015} Birnstiel, T., Andrews, S.~M., Pinilla, P., \& Kama, M.\ 2015, \apjl, 813, L14 

\bibitem[Brice{\~n}o et al.(2002)]{briceno2002} Brice{\~n}o, C., Luhman, K.~L., Hartmann, L., Stauffer, J.~R., \& Kirkpatrick, J.~D.\ 2002, \apj, 580, 317 

\bibitem[Chauvin et al.(2004)]{chauvin2004} Chauvin, G., Lagrange, A.-M., Dumas, C., et al.\ 2004, \aap, 425, L29 

\bibitem[Crida et al.(2006)]{crida2006} Crida, A., Morbidelli, A., \& Masset, F.\ 2006, \icarus, 181, 587 

\bibitem[Daemgen et al.(2016)]{daemgen2016} Daemgen, S., Natta, A., Scholz, A., et al.\ 2016, \aap, 594, A83 

\bibitem[de Boer et al.(2016)]{boer2016} de Boer, J., Salter, G., Benisty, M., et al.\ 2016, \aap, 595, A114 

\bibitem[de Juan Ovelar et al.(2016)]{ovelar2016} de Juan Ovelar, M., Pinilla, P., Min, M., Dominik, C., \& Birnstiel, T.\ 2016, \mnras, 459, L85 

\bibitem[Dittrich et al.(2013)]{dittrich2013} Dittrich, K., Klahr, H., \& Johansen, A.\ 2013, \apj, 763, 117 

\bibitem[Draine(2006)]{draine2006} Draine, B.~T.\ 2006, \apj, 636, 1114 

\bibitem[Espaillat et al.(2014)]{espaillat2014} Espaillat, C., Muzerolle, J., Najita, J., et al.\ 2014, Protostars and Planets VI, 497 

\bibitem[Fedele et al.(2017)]{fedele2017} Fedele, D., Carney, M., Hogerheijde, M.~R., et al.\ 2017, \aap, 600, A72 

\bibitem[Flaherty et al.(2015)]{flaherty2015} Flaherty, K.~M., Hughes, A.~M., Rosenfeld, K.~A., et al.\ 2015, \apj, 813, 99 

\bibitem[Flock et al.(2015)]{flock2015} Flock, M., Ruge, J.~P., Dzyurkevich, N., et al.\ 2015, \aap, 574, A68 

\bibitem[Flock et al.(2016)]{flock2016} Flock, M., Fromang, S., Turner, N.~J., \& Benisty, M.\ 2016, \apj, 827, 144 

\bibitem[Fromang \& Nelson(2006)]{fromang2006} Fromang, S., \& Nelson, R.~P.\ 2006, \aap, 457, 343 

\bibitem[Furlan et al.(2011)]{furlan2011} Furlan, E., Luhman, K.~L., Espaillat, C., et al.\ 2011, \apjs, 195, 3 

\bibitem[Ginski et al.(2016)]{ginski2016} Ginski, C., Stolker, T., Pinilla, P., et al.\ 2016, \aap, 595, A112 

\bibitem[Gundlach \& Blum(2015)]{gundlach2015} Gundlach, B., \& Blum, J.\ 2015, \apj, 798, 34 

\bibitem[Harvey et al.(2012a)]{harvey2012a} Harvey, P.~M., Henning, T., M{\'e}nard, F., et al.\ 2012a, \apjl, 744, L1 

\bibitem[Harvey et al.(2012b)]{harvey2012b} Harvey, P.~M., Henning, T., Liu, Y., et al.\ 2012b, \apj, 755, 67 

\bibitem[Hartmann et al.(1998)]{hartmann1998} Hartmann, L., Calvet, N., Gullbring, E., \& D'Alessio, P.\ 1998, \apj, 495, 385 

\bibitem[Henning et al.(1995)]{henning1995} Henning, T., Michel, B., \& Stognienko, R.\ 1995, \planss, 43, 1333 

\bibitem[Hendler et al.(2017)]{hendler2017} Hendler, N., Mulders, G.~D., Pascucci, I., et al.\ 2017, arXiv:1705.01952 

\bibitem[Herczeg et al.(2009)]{herczeg2009} Herczeg, G.~J., Cruz, K.~L., \& Hillenbrand, L.~A.\ 2009, \apj, 696, 1589 

\bibitem[Hildebrand(1983)]{hildebrand1983} Hildebrand, R.~H.\ 1983, \qjras, 24, 267 

\bibitem[Ilgner \& Nelson(2006)]{ilgner2006} Ilgner, M., \& Nelson, R.~P.\ 2006, \aap, 455, 731 

\bibitem[Isella et al.(2016)]{isella2016} Isella, A., Guidi, G., Testi, L., et al.\ 2016, Phys. Rev. Lett., 117, 25

\bibitem[Joergens et al.(2012)]{joergens2012} Joergens, V., Pohl, A., Sicilia-Aguilar, A., \& Henning, T.\ 2012, \aap, 543, A151 

\bibitem[Johnson et al.(2012)]{johnson2012} Johnson, J.~A., Gazak, J.~Z., Apps, K., et al.\ 2012, \aj, 143, 111 

\bibitem[Kataoka et al.(2013)]{kataoka2013} Kataoka, A., Tanaka, H., Okuzumi, S., \& Wada, K.\ 2013, \aap, 557, L4 

\bibitem[Kataoka et al.(2014)]{kataoka2014} Kataoka, A., Okuzumi, S., Tanaka, H., \& Nomura, H.\ 2014, \aap, 568, A42 

\bibitem[Kenyon \& Hartmann(1987)]{kenyon1987} Kenyon, S.~J., \& Hartmann, L.\ 1987, \apj, 323, 714 

\bibitem[Klahr \& Henning(1997)]{klahr1997} Klahr, H.~H., \& Henning, T.\ 1997, \icarus, 128, 213 

\bibitem[Klein et al.(2003)]{klein2003} Klein, R., Apai, D., Pascucci, I., Henning, T., \& Waters, L.~B.~F.~M.\ 2003, \apjl, 593, L57 

\bibitem[Liu et al.(2015)]{liu2015} Liu, Y., Joergens, V., Bayo, A., Nielbock, M., \& Wang, H.\ 2015, \aap, 582, A22 

\bibitem[Lodato et al.(2005)]{lodato2005} Lodato, G., Delgado-Donate, E., \& Clarke, C.~J.\ 2005, \mnras, 364, L91 

\bibitem[Luhman et al.(2003)]{luhman2003} Luhman, K.~L., Brice{\~n}o, C., Stauffer, J.~R., et al.\ 2003, \apj, 590, 348 

\bibitem[Luhman(2004)]{luhman2004} Luhman, K.~L.\ 2004, \apj, 617, 1216 

\bibitem[Luhman et al.(2007)]{luhman2007} Luhman, K.~L., Adame, L., D'Alessio, P., et al.\ 2007, \apj, 666, 1219 

\bibitem[Mart{\'{\i}}n et al.(2001)]{martin2001} Mart{\'{\i}}n, E.~L., Dougados, C., Magnier, E., et al.\ 2001, \apjl, 561, L195 

\bibitem[Masset(2000)]{masset2000} Masset, F.\ 2000, \aaps, 141, 165 

\bibitem[Mulders \& Dominik(2012)]{mulders2012} Mulders, G.~D., \& Dominik, C.\ 2012, \aap, 539, A9 

\bibitem[Muzerolle et al.(2006)]{muzerolle2006} Muzerolle, J., Adame, L., D'Alessio, P., et al.\ 2006, \apj, 643, 1003 

\bibitem[Muzerolle et al.(2010)]{muzerolle2010} Muzerolle, J., Allen, L.~E., Megeath, S.~T., Hern{\'a}ndez, J., \& Gutermuth, R.~A.\ 2010, \apj, 708, 1107 

\bibitem[Owen et al.(2011)]{owen2011} Owen, J.~E., Ercolano, B., \& Clarke, C.~J.\ 2011, \mnras, 412, 13 

\bibitem[Owen(2016)]{owen2016} Owen, J.~E.\ 2016, \pasa, 33, e005 

\bibitem[Padoan \& Nordlund(2004)]{padoan2004} Padoan, P., \& Nordlund, AA \ 2004, \apj, 617, 559 

\bibitem[Palau et al.(2014)]{palau2014} Palau, A., Zapata, L.~A., Rodr{\'{\i}}guez, L.~F., et al.\ 2014, \mnras, 444, 833 

\bibitem[Pascucci et al.(2009)]{pascucci2009} Pascucci, I., Apai, D., Luhman, K., et al.\ 2009, \apj, 696, 143 

\bibitem[Pascucci et al.(2016)]{pascucci2016} Pascucci, I., Testi, L., Herczeg, G.~J., et al.\ 2016, \apj, 831, 125 

\bibitem[Paszun \& Dominik(2006)]{paszun2006} Paszun, D., \& Dominik, C.\ 2006, \icarus, 182, 274

\bibitem[Payne \& Lodato(2007)]{payne2007} Payne, M.~J., \& Lodato, G.\ 2007, \mnras, 381, 1597 

\bibitem[Pinilla et al.(2012)]{pinilla2012} Pinilla, P., Benisty, M., \& Birnstiel, T.\ 2012, A\&A, 545, A81

\bibitem[Pinilla et al.(2013)]{pinilla2013} Pinilla, P., Birnstiel, T., Benisty, M., et al.\ 2013, \aap, 554, A95 

\bibitem[Pinilla et al.(2014)]{pinilla2014} Pinilla, P., Benisty, M., Birnstiel, T., et al.\ 2014, \aap, 564, A51 

\bibitem[Pinilla et al.(2015)]{pinilla2015} Pinilla, P., de Juan Ovelar, M., Ataiee, S., et al.\ 2015, \aap, 573, A9 

\bibitem[Pinilla et al.(2016)]{pinilla2016} Pinilla, P., Flock, M., Ovelar, M.~d.~J., \& Birnstiel, T.\ 2016, \aap, 596, A81 

\bibitem[Reipurth \& Clarke(2001)]{reipurth2001} Reipurth, B., \& Clarke, C.\ 2001, \aj, 122, 432 

\bibitem[Ricci et al.(2010)]{ricci2010} Ricci, L., Testi, L., Natta, A., et al.\ 2010, \aap, 512, A15 

\bibitem[Ricci et al.(2012a)]{ricci2012} Ricci, L., Testi, L., Natta, A., Scholz, A., \& de Gregorio-Monsalvo, I.\ 2012a, \apjl, 761, L20 

\bibitem[Ricci et al.(2012b)]{ricci2012b} Ricci, L., Trotta, F., Testi, L., et al.\ 2012b, \aap, 540, A6 

\bibitem[Ricci et al.(2013)]{ricci2013} Ricci, L., Isella, A., Carpenter, J.~M., \& Testi, L.\ 2013, \apjl, 764, L27 

\bibitem[Ricci et al.(2014)]{ricci2014} Ricci, L., Testi, L., Natta, A., et al.\ 2014, \apj, 791, 20 

\bibitem[Rigliaco et al.(2011)]{rigliaco2011} Rigliaco, E., Natta, A., Randich, S., et al.\ 2011, \aap, 526, L6 

\bibitem[Shakura \& Sunyaev(1973)]{shakura1973} Shakura, N.~I., \& Sunyaev, R.~A.\ 1973,A\&A, 24, 337  

\bibitem[Simon \& Armitage(2014)]{simon2014} Simon, J.~B., \& Armitage, P.~J.\ 2014, \apj, 784, 15 

\bibitem[Schmidt et al.(2014)]{schmidt2014} Schmidt, S.~J., Prieto, J.~L., Stanek, K.~Z., et al.\ 2014, \apjl, 781, L24 

\bibitem[Scholz et al.(2006)]{scholz2006} Scholz, A., Jayawardhana, R., \& Wood, K.\ 2006, \apj, 645, 1498 

\bibitem[Sz{\H u}cs et al.(2010)]{szucs2010} Sz{\H u}cs, L., Apai, D., Pascucci, I., \& Dullemond, C.~P.\ 2010, \apj, 720, 1668 

\bibitem[Strom et al.(1989)]{strom1989} Strom, K.~M., Strom, 
S.~E., Edwards, S., Cabrit, S., \& Skrutskie, M.~F.\ 1989, AJ, 97, 1451

\bibitem[Teague et al.(2016)]{teague2016} Teague, R., Guilloteau, S., Semenov, D., et al.\ 2016, \aap, 592, A49 

\bibitem[Testi et al.(2014)]{testi2014} Testi, L., Birnstiel, T., Ricci, L., et al.\ 2014, Protostars and Planets VI, 339 

\bibitem[Testi et al.(2016)]{testi2016} Testi, L., Natta, A., Scholz, A., et al.\ 2016, \aap, 593, A111 

\bibitem[Uribe et al.(2011)]{uribe2011} Uribe, A.~L., Klahr, H., Flock, M., \& Henning, T.\ 2011, \apj, 736, 85 

\bibitem[van Boekel et al.(2017)]{boekel2017} van Boekel, R., Henning, T., Menu, J., et al.\ 2017, \apj, 837, 132 

\bibitem[van der Plas et al.(2016)]{vanderplas2016} van der Plas, G., M{\'e}nard, F., Ward-Duong, K., et al.\ 2016, \apj, 819, 102 

\bibitem[van der Plas et al.(2017)]{vanderplas2017} van der Plas, G., Wright, C.~M., M{\'e}nard, F., et al.\ 2017, \aap, 597, A32 

\bibitem[Wada et al.(2009)]{wada2009} Wada, K., Tanaka, H., Suyama, T., Kimura, H., \& Yamamoto, T.\ 2009, ApJ, 702, 1490

\bibitem[Wada et al.(2011)]{wada2011} Wada, K., Tanaka, H., Suyama, T., Kimura, H., \& Yamamoto, T.\ 2011, ApJ, 737, 36 

\bibitem[Ward-Duong et al.(submitted)]{ward2017}  Ward-Duong, K., Patience, J., Bulger, J, et al.\ 2017, submitted to AJ.

\bibitem[Weidenschilling(1977)]{weidenschilling1977} Weidenschilling, S.~J.\ 1977, \mnras, 180, 57

\bibitem[Whipple(1972)]{whipple1972} Whipple, F.~L. 1972, From Plasma to Planet, 211

\bibitem[Wichmann et al.(1998)]{wichmann1998} Wichmann, R., Bastian, U., Krautter, J., Jankovics, I., \& Rucinski, S.~M.\ 1998, \mnras, 301, L39 

\end{thebibliography}
\end{document}